\documentclass[prl,superscriptaddress,reprint,showpacs,longbibliography]{revtex4-2}
\usepackage{lineno}
\usepackage{color}
\usepackage{physics}
\usepackage{graphicx,textcomp,amssymb,amsmath,dcolumn,hyperref}
\usepackage{siunitx}
\usepackage{comment}
\newcommand {\bkt} [1] {\langle #1 \rangle}
\usepackage{romannum}
\newcommand{\prob}{\mbox{prob}}
\newcommand{\pdf}{\mbox{pdf}}
\begin{document}

\title{
Long-lived valley states in bilayer graphene quantum dots
}

\author{Rebekka Garreis}
\email{garreisr@phys.ethz.ch}
\email{ctong@phys.ethz.ch}
\thanks{\\ These authors contributed equally to this paper.}

\author{Chuyao Tong}
\email{garreisr@phys.ethz.ch}
\email{ctong@phys.ethz.ch}
\thanks{\\ These authors contributed equally to this paper.}

\author{Jocelyn Terle}

\author{Max Josef Ruckriegel} 
\author{Jonas Daniel Gerber} 
\author{Lisa Maria Gächter} 

\affiliation{Laboratory for Solid State Physics, ETH Zurich, CH-8093 Zurich, Switzerland}

\author{Kenji Watanabe}
\affiliation{Research Center for Functional Materials, National Institute for Materials Science, 1-1 Namiki, Tsukuba 305-0044, Japan}
\author{Takashi Taniguchi}
\affiliation{International Center for Materials Nanoarchitectonics, National Institute for Materials Science,  1-1 Namiki, Tsukuba 305-0044, Japan}

\author{Thomas Ihn}
\author{Klaus Ensslin}
\author{Wei Wister Huang}

\affiliation{Laboratory for Solid State Physics, ETH Zurich, CH-8093 Zurich, Switzerland}


\begin{abstract}
{\bf Bilayer graphene is a promising platform for electrically controllable qubits in a two-dimensional material. Of particular interest is the ability to encode quantum information in the so-called valley degree of freedom, a two-fold orbital degeneracy that arises from the symmetry of the hexagonal crystal structure. The use of valleys could be advantageous, as known spin- and orbital-mixing mechanisms are unlikely to be at work for valleys, promising more robust qubits.
The Berry curvature associated with valley states allows for electrical control of their energies, suggesting routes for coherent qubit manipulation. However, the relaxation time of valley states — which ultimately limits these qubits’ coherence properties and therefore their suitability as practical qubits — is not yet known. Here, we measure the characteristic relaxation times of these spin and valley states in gate-defined bilayer graphene quantum dot devices. Different valley states can be distinguished from each other with a fidelity of over \SI{99}{\percent}. The relaxation time between valley triplets and singlets exceeds 500ms, and is more than one order of magnitude longer than for spin states. This work facilitates future measurements on valley-qubit coherence, demonstrating bilayer graphene as a practical platform hosting electrically controlled long-lived valley qubits.}

\end{abstract}
\maketitle

Bilayer graphene (BLG) offers unique opportunities as a host material for spin qubits~\cite{Trauzettel2007spinqubit,RMPspin}. These include weak spin--orbit interactions~\cite{Kurzmann2021,Banszerus2022} and natural nuclear-spin concentrations as low as 1.1$\%$ (compared to 4.7$\%$ in Si), which can be further improved by isotopic purification~\cite{Chen2012Thermal}. Moreover, 2D materials allow for the realisation of smaller transistors~\cite{Liu2021} and possibly more strongly coupled quantum devices, as compared to bulk materials. 

In addition, the symmetry of the hexagonal Bravais lattice of BLG gives rise to a valley degeneracy, which behaves analogously to spins~\cite{Eich2018, Tong2021, Garreis2021, Recher2009}.
This unique valley degeneracy in BLG with electrically tunable valley $g$-factor~\cite{Tong2021} provides an additional degree of freedom to realise and manipulate qubits. In particular, there is the prospect of realising highly robust qubits with valley states. That is, whereas charge qubits couple to electric fields and spin qubits to magnetic fields, valley qubits consist of two degenerate states with the same charge distribution and the same spin configuration, but differ in their locations in reciprocal space. Theories have proposed various intervalley scattering mechanisms, requiring a short-range event on the scale of the lattice period~\cite{Palyi2009, Morpurgo2006}. Hence, for sufficiently low atomic defect density, the valley lifetimes are expected to be limited not by intrinsic mixing mechanisms such as phonon-mediated spin--valley coupling, but rather by the finite size of the dot ultimately breaking translational invariance, similar as it has been discussed for transition metal dichalcogenides~\cite{Liu2014}. However, in optically addressed valley qubits in other 2D materials, so far only very short valley life times have been measured~\cite{Soni2022}.

The development of BLG quantum dot devices have made rapid progresses in recent years~\cite{Eich2018, Banszerus2018, Eich2018doubledot, Banszerus2020singleelectron, Eich2020}, with the demonstration of high quality and controllability~\cite{Banszerus2020, Tong2021} and the discovery of intriguing physics~\cite{Kurzmann219excited, Kurzmann2021, Garreis2021, Banszerus2022, Tong2022spinmixing}, such as switchable Pauli spin- and valley-blockade in coupled double quantum dots~\cite{Tong2022}, as well as the realization of high-quality charge sensing technology~\cite{Kurzmann2019Detector,Garreis2022}. In recent experiments we found spin-relaxation times $T_1$ of up to $\SI{50}{ms}$ measured with the single-shot Elzerman readout technique~\cite{Elzerman2004} in a single quantum 
dot~\cite{GachterGarreis2022},
comparable with values from other semiconductor quantum dot systems~\cite{stano2021review}, such as in \Romannum{3}-\Romannum{5}~\cite{Nakajima2020,Cerfontaine2014,Nichol2017}, silicon-~\cite{xue2021computing,Zajac2018,Yoneda2018,mills2021twoqubit} and germanium-~\cite{Hendrickx2021} based heterostructures. 
Here we demonstrate single-shot readout with both spin and valley Pauli blockade~\cite{johnson2005triplet,Zheng2019} in gate-defined BLG double quantum dots, and thereby the measurement of characteristic spin and valley relaxation times $T_1$ between spin- or valley-triplet and singlet states. Unique to BLG, we can select between spin- or valley-blockade regimes by choosing appropriate perpendicular magnetic fields~\cite{Tong2022}. The spin-$T_1$ is measured to be up to $\SI{60}{ms}$ at $B_\mathrm{\perp}=\SI{700}{mT}$, corroborating our recent findings in single quantum dots~\cite{GachterGarreis2022}. By increasing the interdot tunnel coupling, the spin $T_1$ time is reduced. Moreover, we observe outstandingly long valley $T_1$ times, longer than $\SI{500}{ms}$ at $B_\mathrm{\perp}=\SI{250}{mT}$. Unlike in the relaxation of spin states, intervalley relaxation times are found to be robust against variation of the interdot tunnel coupling strength. This valley lifetime is comparable with the state-of-the-art spin singlet-triplet $T_1$ measured in Si/SiGe and Si/SiO$_2$, and an order of magnitude longer than their $T_1$ reported at such low magnetic field~\cite{PhysRevLett.108.046808,yang2020operation}. 

\begin{figure}
    \includegraphics[width=8.5cm]{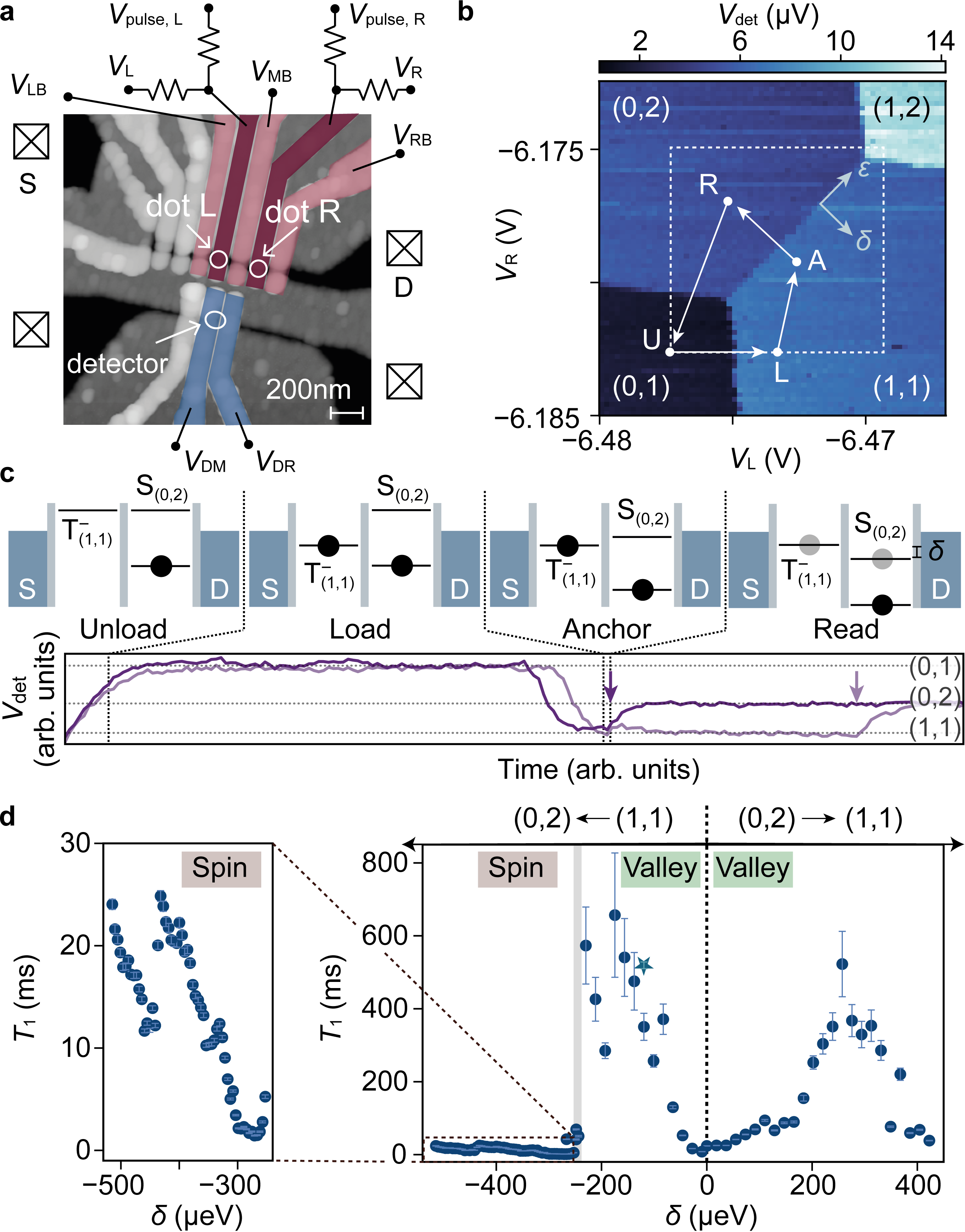}
	\caption{Pulse protocol used to determine spin and valley relaxation times. \textbf{a} False-colour atomic force microscopic image of the device. The double quantum dot is formed underneath gates L and R (dark red), with tunable tunnel barrier gates (light red). In the other channel, a detector quantum dot with plunger gates DM and DR is capacitively coupled to the double quantum dots and serves as a charge sensor. 
    \textbf{b} Charge-stability map probed with detector signal $V_\mathrm{det}$ measured across the current-biased detector channel, around the $(1,1)$---$(0,2)$ charge transition. Axes of total energy $\epsilon$ and energy detuning $\delta$ are marked. The dashed square with a side length of \SI{8}{mV} marks the experimentally accessible pulse window, limited by the amplitude of the pulse generator.
    \textbf{c} Spin and valley pulsing protocol for $(1,1)\rightarrow(0,2)$ with corresponding exemplary time traces. The gate position for each phase is marked in panel \textbf{b}. Starting from the $(0,1)$ state in unload (U), we pulse to the load position (L) slower than the dot-lead tunnel rate to prepare a $(1,1)$ state. We then pulse to the anchor point (A) slower than the line-bandwidth of the order of $\SI{100}{\micro s}$, before pulsing to the read location (R). In the read phase, the system is energetically preferred to transfer to $(0,2)$, but is forbidden to do so by mismatching quantum numbers between triplet $(1,1)\mathrm{T^-}$ and singlet $(0,2)\mathrm{S}$, unless a relaxation event occurs (indicated by purple arrows in the time traces). 
    \textbf{d} Characteristic spin and valley relaxation times $T_1$ measured at perpendicular magnetic field $B_\mathrm{\perp}=\SI{250}{mT}$ along the $\delta$-axis
    . Valley $T_1$ values exceeding $\SI{500}{ms}$ are measured, much longer than the measured spin $T_1\sim\SI{10}{ms}$. Error bars correspond to the standard deviation of the calculated $T_1$. The star marks measurements with read time of $\SI{1.5}{s}$.}
	\label{fig1}
\end{figure}

\begin{figure*}
    \includegraphics[width=17.8cm]{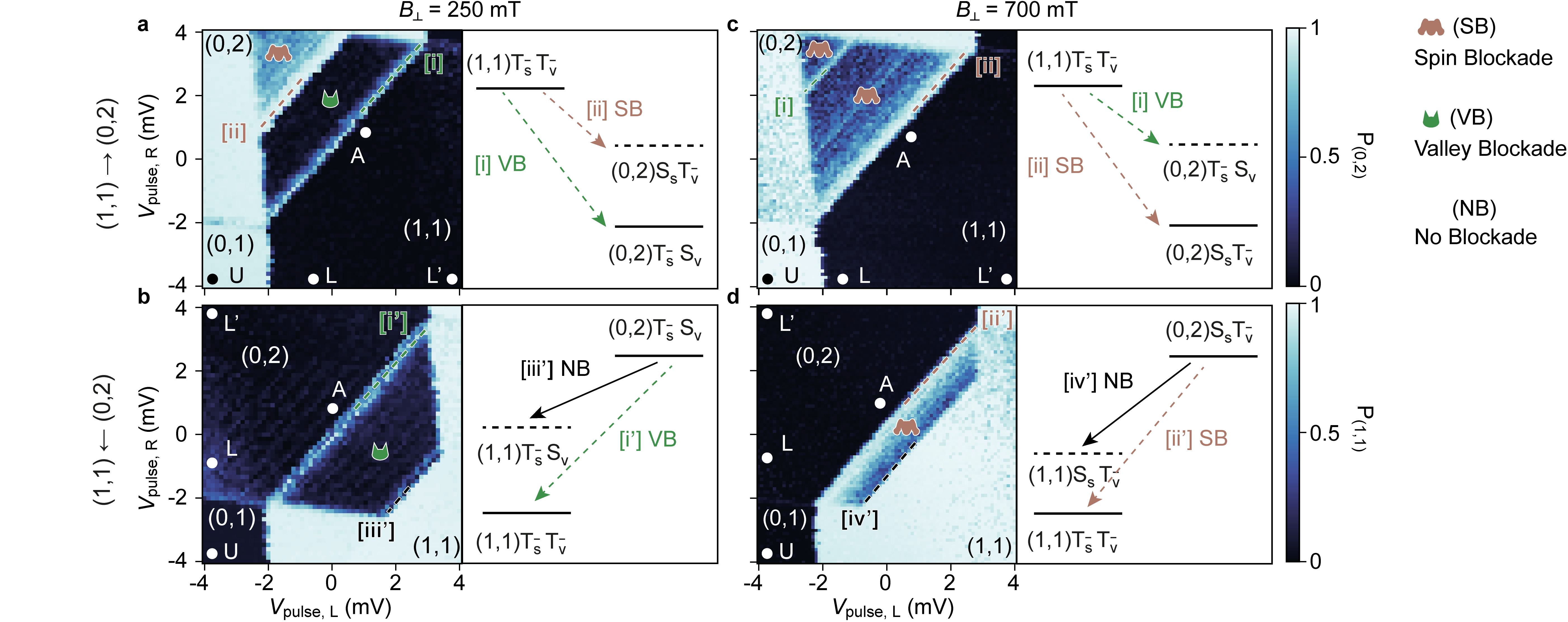}
	\caption{Probability of occupation in \textbf{a,c} $(0,2)$, and \textbf{b,d} $(1,1)$ states during the read phase, and schematics showing the relevant states and transitions involved, for pulsing $(1,1)\rightarrow(0,2)$, and $(0,2)\rightarrow(1,1)$ while loading \emph{and} anchoring at locations L', at $B_\mathrm{\perp}=$ \textbf{a,b} $\SI{250}{mT}$, and \textbf{c,d} $\SI{700}{mT}$. Green and brown dots label regions of valley and spin blockade, respectively, shown by the low occupation probability (dark) of the target states. Dashed lines indicate relevant state alignments. For $(1,1)\rightarrow(0,2)$ at $B_\mathrm{\perp}=\SI{250}{mT}$, the strong valley blockade from $(1,1)\mathrm{T^-_sT^-_v}$ to the valley-mismatched ground state $(0,2)\mathrm{T^-_sS_v}$ is replaced by the weaker spin blockade to the spin-mismatched excited state $(0,2)\mathrm{S_sT^-_v}$ at [ii]; while reversely at $B_\mathrm{\perp}=\SI{700}{mT}$, the ground-state spin blockade cannot be lifted by the stronger excited-state valley blockade at [i]. For the rest of the measurements, loading is performed at location L and anchoring at A.}
	\label{FIG2}
\end{figure*}

The BLG double quantum dots studied here are defined by electrostatic gating in the sample shown in Fig.~\ref{fig1}a (for details on sample fabrication and quantum dot tuning, see Methods). In one channel, we define the two quantum dots L and R underneath their respective plunger gates (dark red) with voltages $V_\mathrm{L}$ and $V_\mathrm{R}$. The dot--lead and interdot tunnel couplings are controlled individually by the barrier gate (light red) voltages $V_\mathrm{LB,RB}$ and $V_\mathrm{MB}$, respectively. Separated by a depletion region, a third quantum dot (labelled `detector' in Fig.~\ref{fig1}a) formed in the neighbouring channel is controlled by plunger-gate (blue) voltages $V_\mathrm{DM}$ and $V_\mathrm{DR}$. This dot serves as a charge sensor, as it is capacitively coupled to dots L and R, more strongly to the left than to the right dot. A change in the double-dot charge configuration constitutes a discrete change of the electrostatic environment of the sensor dot, thereby giving rise to a step in $V_\mathrm{det}$, the voltage measured across the detection channel when applying a constant current bias~\cite{Kurzmann2019Detector}. The sensor dot is tuned to be at the rising or falling edge of a conductance resonance for optimised sensitivity and detection bandwidth~\cite{Garreis2022}.

We tune the double dot to the previously studied~\cite{Tong2022} two-electron configuration near the $(1,1)$--$(0,2)$ charge degeneracy, where $(N_\mathrm{L}, N_\mathrm{R})$ labels the number of electrons in the left and in the right dot. With long integration time ($\SI{20}{ms}$) in the detector circuit, the relevant charge states manifest themselves as discrete values of $V_\mathrm{det}$, as shown in the charge stability map (Fig.~\ref{fig1}b). All four charge states $(1,1)$, $(0,2)$, $(0,1)$ and $(1,2)$ can be clearly distinguished. For single-shot readout, we fix $V_\mathrm{L}$ and $V_\mathrm{R}$ to an operating point, and apply voltage pulses $V_\mathrm{pulse,L}$, and $V_\mathrm{pulse,R}$ to the left and the right plunger gates (see Fig.~\ref{fig1}a for the schematic circuit). We collect real-time data during the pulsed experiments at a sampling rate of \SI{27.5}{kHz}.

Figures~\ref{fig1}b,c depict the readout protocol for measuring inelastic relaxations between $(1,1)$ and $(0,2)$ states. Before describing which quantum states are addressed in the two charge configurations, we first discuss this protocol as an example to explain our general measurement scheme. Starting at the unload point U with a $(0,1)$ state, we first prepare a $(1,1)$ state by pulsing from U to the load (L) configuration within time $t_\mathrm{L}$, which is much longer than the dot--lead tunnelling time, so that an electron can tunnel into the left dot with high fidelity while the $(0,2)$ level remains above the Fermi energy of the leads.
We then pulse to an anchor point (A), where both the $(1,1)$ and the $(0,2)$ states are well below the Fermi energy in the leads, with $(1,1)$ remaining in the ground state. Subsequently, we pulse quickly (bandwidth-limited to the order of $\SI{10}{kHz}$) to the read position (R), where the $(0,2)$ state is lower in energy than $(1,1)$. The electron in the left dot is now energetically allowed to relax into the right dot while releasing its excess energy into the environment. We wait at this position for a time $t_\mathrm{R}$ before returning to the unload point U. An inelastic transition between the $(1,1)$ and the $(0,2)$ state happening within the time-interval $t_\mathrm{R}$ is detected in real time. Figure~\ref{fig1}c shows two exemplary time traces with the inelastic transitions occurring at different times (purple arrows). Repeating this pulse sequence at least $\num{10000}$ times, we obtain the statistical distribution of the inelastic relaxation times, from which we extract the average relaxation time $T_1$. A similar scheme with points L and A in the $(0,2)$ and R in the $(1,1)$ region is applied for pulsing from an initial $(0,2)$ into the $(1,1)$ charge configuration for measuring the $T_1$ time of $(0,2)\rightarrow(1,1)$ transitions.

In order to select the specific quantum states between which we wish to measure the $T_1$ time, we apply a magnetic field perpendicular to the graphene plane~\cite{Kurzmann219excited, Tong2022, Moeller2021, Knothe2020} and study the relaxation between Pauli-blockaded valley and spin states at $B_\perp=\SI{250}{mT}$ and $\SI{700}{mT}$, respectively. In both cases, the lowest-energy $(1,1)$ states are valley- and spin-polarised (valley-triplet $\mathrm{T^-_v}$ and spin-triplet $\mathrm{T^-_s}$ state).

At $B_\perp=\SI{250}{mT}$, the $(0,2)$ ground state is a valley-singlet $\mathrm{S_v}$ spin-triplet state $\mathrm{T^-_s}$; 
the $(1,1)\leftrightarrow(0,2)$ transitions are therefore valley-blockaded. By contrast, at $B_\perp=\SI{700}{mT}$, the $(0,2)$ ground state has turned into a valley-polarised (valley-triplet $\mathrm{T^-_v}$) spin-singlet state, such that the $(1,1)\leftrightarrow (0,2)$ transitions are spin-blockaded.

Figure~\ref{fig1}d shows the main result of this paper for the $T_1$ times measured at $B_{\perp}=\SI{250}{mT}$ as a function of detuning $\delta$ of the read position R from the charge transition line (orientation of $\delta$-axis marked in Fig.~\ref{fig1}b).
At sufficiently small detuning $\SI{-250}{\micro eV}<\delta<\SI{420}{\micro eV}$, a valley flip is required to lift the valley blockade, so that the resulting $T_1$ times can be identified with the valley relaxation time. We observe exceptionally long relaxation times of $T_1>\SI{500}{ms}$, demonstrating that the valley states are remarkably long-lived in this double quantum dot system. This suggests that valley flips are suppressed within the dot and also during tunnelling, indicating that valley states are highly suitable for qubit operation. At $\delta<\SI{-250}{\micro eV}$, the $(0,2)$ excited state with valley-triplet spin-singlet character is lower in energy than the $(1,1)$ ground state and the valley-blockade can therefore be circumvented with a spin-flip transition to this state. In this regime, we measure the spin-relaxation time $T_1\leq \SI{25}{ms}$, which is by an order of magnitude shorter than the valley relaxation time, but still comparable with values observed in other semiconductor quantum dot systems, and sufficiently long for high-fidelity qubit operation and readout.

As an illustration of the details that lead to and go beyond the results presented in Fig.~\ref{fig1}d, we show data obtained from pulse cycles at $B_\mathrm{\perp}=$ $\SI{250}{mT}$ and $\SI{700}{mT}$ in Figs.~\ref{FIG2}. We pulsed during the load phase from U deep into i the $(1,1)$ or ii the $(0,2)$ configurations at locations L' (marked in Fig.~\ref{FIG2}) loading predominantly the respective ground states. Then, fixing U and L', and anchoring at L', we raster-scanned the read configuration R over the region marked with the dashed square in Fig.~\ref{fig1}b, and repeated at least $50$ pulse cycles for each point. The $V_\mathrm{det}$ signal averaged over the read time $t_\mathrm{R}$ and over all the pulse cycles reflects the probabilities $P_\mathrm{(0,2)}$ (Fig.~\ref{FIG2}a,c) and $P_\mathrm{(1,1)}$ (Fig.~\ref{FIG2}b,d) at R~\cite{johnson2005triplet, Churchill2009}. The resulting normalised probability maps are shown in Fig.~\ref{FIG2} for a,b $(1,1)\rightarrow(0,2)$, and c,d $(0,2)\rightarrow(1,1)$. For a detailed description of the relevant $(1,1)$ and $(0,2)$ states involved and their evolution in the magnetic field $B_z$, see Supplementary Information~A. 

At $B_\mathrm{\perp}=\SI{250}{mT}$ (Figs~\ref{fig1}d and \ref{FIG2}a,b), the $(0,2)$ ground state is the valley-singlet spin-triplet $\mathrm{T^-_sS_v}$. The bundle of $(1,1)$ states (containing all spin-triplet and -singlet states, split-off by the Zeeman energy and $\Delta_\mathrm{SO}$) with polarised valleys $\mathrm{T^-_v}$ is lower in energy than the bundle of $(1,1)\mathrm{S_v/T^0_v}$ by $g_\mathrm{v}\mu_\mathrm{B}B_\mathrm{\perp}$, where $\mu_\mathrm{B}$ is the Bohr magneton and $g_\mathrm{v}$, approximately ${20}$, the dot-geometry-dependent valley $g$-factor, as the energies of the valley states couple to a perpendicular magnetic field, similar to the Zeeman effect for spins. Therefore, in Fig.~\ref{FIG2}a, a strongly blockaded region (green) for $(1,1)\rightarrow(0,2)$ is observed, where the system remains mostly in $(1,1)\mathrm{T^-_v}$ during reading ($\SI{26}{ms}$), as its transition [i] to the ground state $(0,2)\mathrm{S_v}$ is valley-blockaded. At large enough $|\delta|$, when the $(0,2)\mathrm{S_sT^-_v}$ excited state becomes accessible, the spin-blockaded transition [ii] to this state can circumvent the valley-blockaded transition [i] at the cost of a spin-flip. As the valley-blockaded region (green) is lifted by spin blockade (brown) on transition [ii] with shorter yet still finite relaxation time, we conclude that spin flips occur more frequently than valley flips. The $(0,2)$ excited state with matching quantum numbers $\mathrm{T^-_sT^-_v}$, able to lift both spin and valley blockade, occurs at much higher energies~\cite{Moeller2021}. A strongly blockaded region (green in Fig.~\ref{FIG2}b) is also observed for $(0,2)\rightarrow(1,1)$, as transition [i'] from the $(0,2)$ ground state $\mathrm{S_v}$ to the $(1,1)$ ground state $\mathrm{T^-_v}$ is valley-blockaded. This blockade is completely lifted at large enough $|\delta|$ at [iii'], giving access to the excited state $(1,1)\mathrm{T^-_sS_v}$ with matching quantum numbers. We observe valley blockade stemming from the same set of states and transitions at even lower magnetic field, as low as at $B_\mathrm{\perp} = \SI{20}{mT}$ (see Supplementary Information~C).

With this scenario in mind, we move to $B_\mathrm{\perp}=\SI{700}{mT}$ (Fig.~\ref{FIG2}c,d), where the $(0,2)$ ground state changes to $(0,2)\mathrm{S_sT^-_v}$, due to $\mathrm{T^-_v}$ being lowered in energy by the coupling of the valleys to $B_\mathrm{\perp}$. The [i] valley- and [ii] spin-blockaded transitions thus reverse their order in energy relative to the scenario at $B_\mathrm{\perp}=\SI{250}{mT}$. In Fig.~\ref{FIG2}c, the transition [ii] from the loaded $(1,1)\mathrm{T^-_sT^-_v}$ to the spin-mismatched ground state $(0,2)\mathrm{S_sT^-_v}$ gives rise to a spin-blockaded region (brown). Resonance [i] at finite $|\delta|$ appears as the now excited valley-mismatched state $(0,2)\mathrm{T^-_sS_v}$ is accessible in energy. However, unlike at $B_\mathrm{\perp}=\SI{250}{mT}$, this excited state does not lift the spin blockade, as valley flips occur more slowly than spin flips. By contrast, the spin-blockaded (brown) region in Fig.~\ref{FIG2}d is much smaller for $(0,2)\rightarrow(1,1)$, as the $(0,2)$ ground state $\mathrm{T^-_sS_v}$ is spin blockaded and cannot proceed to the ground state $(1,1)\mathrm{T^-_sT^-_v}$ until access to the excited state $(1,1)\mathrm{S_sT^-_v}$ lifts the spin blockade completely at [iv'], $g_\mathrm{s}\mu_\mathrm{B}B_\mathrm{\perp}+\Delta_\mathrm{SO}$ away in $|\delta|$, where $g_\mathrm{s}=2$ is the spin $g$-factor and $\Delta_\mathrm{SO}\sim\SI{60}{\micro eV}$ the zero-field Kane--Mele splitting~\cite{Kurzmann2021,Banszerus2022}. 

\begin{figure}
	\includegraphics[width=8.5cm]{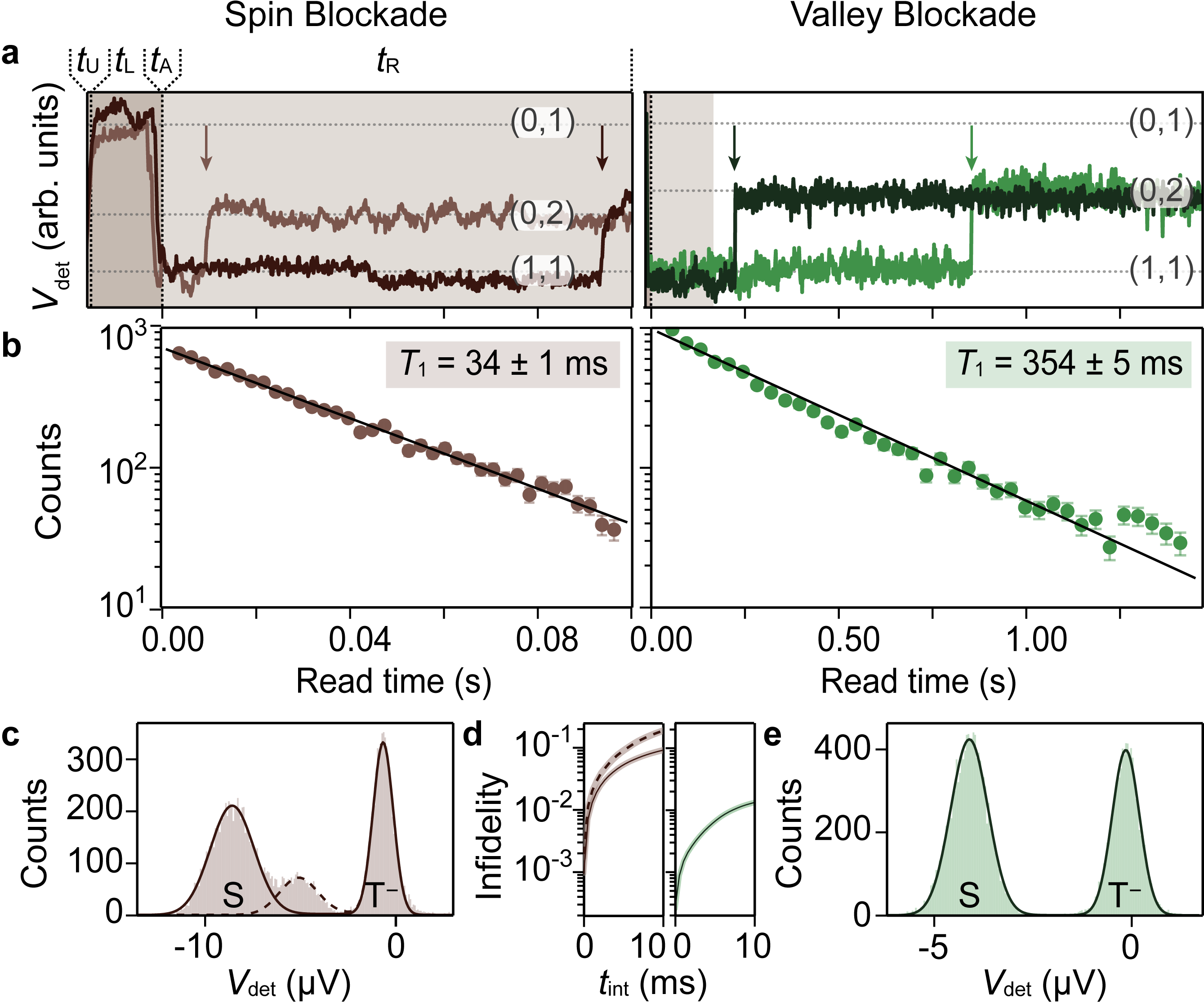}
	\caption{Extracting characteristic relaxation times $T_1$ for $(1,1)\mathrm{T}\rightarrow(0,2)\mathrm{S}$ at $B_\mathrm{\perp}=\SI{700}{mT}$ for spins, and at $B_\mathrm{\perp}=\SI{250}{mT}$ for valleys, for detuning values marked on Fig.~\ref{figcom} by arrows, while loading at L marked in Fig.~\ref{FIG2} to ensure loading of the $(1,1)$ ground state, and anchoring at A. \textbf{a} Exemplary time traces with arrows marking relaxation events for spins on the left and for valleys on the right. \textbf{b} Distribution of relaxation times for \num{10000} pulse cycles, with $T_1$ calculated by solving for the ansatz $e^{-t/T_1}$ with average relaxation time $\bkt{t}$. Error bars correspond to the standard deviation of the binomial distribution of the number of counts. \textbf{c,e} Distribution of $V_\mathrm{det}$ during the first $\SI{1}{ms}$ of the read phase. The (1,1)$\mathrm{T^-}$ probability at the beginning of the read phase is roughly 50\%. To describe the distribution we follow Ref.~\onlinecite{Barthel2009} to include the finite relaxation time during the read phase. For spins we evaluate the two singlet peaks individually with $\mu_{\mathrm{S},1}=\SI{-8.54}{\micro V}$, $\sigma_{\mathrm{S},1}=\SI{1.14}{\micro V}$, $\mu_{\mathrm{S},2}=\SI{-5.07}{\micro V}$, $\sigma_{\mathrm{S},2}=\SI{0.99}{\micro V}$  and $\mu_{\mathrm{T^-},1}=\SI{-0.64}{\micro V}$, $\sigma_{\mathrm{T^-},1}=\SI{0.53}{\micro V}$ and $T_1 = \SI{34}{ms}$, $\tau_M = \SI{1}{ms}$. For valleys, we find $\mu_{\mathrm{S}}=\SI{-4.11}{\micro V}$, $\sigma_{\mathrm{S}}=\SI{0.47}{\micro V}$ and $\mu_{\mathrm{T^-}}=\SI{-0.15}{\micro V}$, $\sigma_{\mathrm{T^-}}=\SI{0.34}{\micro V}$ and $T_1 = \SI{354}{ms}$, $\tau_M = \SI{1}{ms}$. \textbf{d} Measured infidelity as a function of the integration time $t_\mathrm{int}$ of the read phase. For long integration times the relaxation lowers the fidelity. We find an overall maximum fidelity of $99.80\%$ for spins (lower bound) and $99.97\%$ for valleys. Shaded regions of each line indicate one standard deviation.}
	\label{figeg}
\end{figure}

\begin{figure}
	\includegraphics[width=8.5cm]{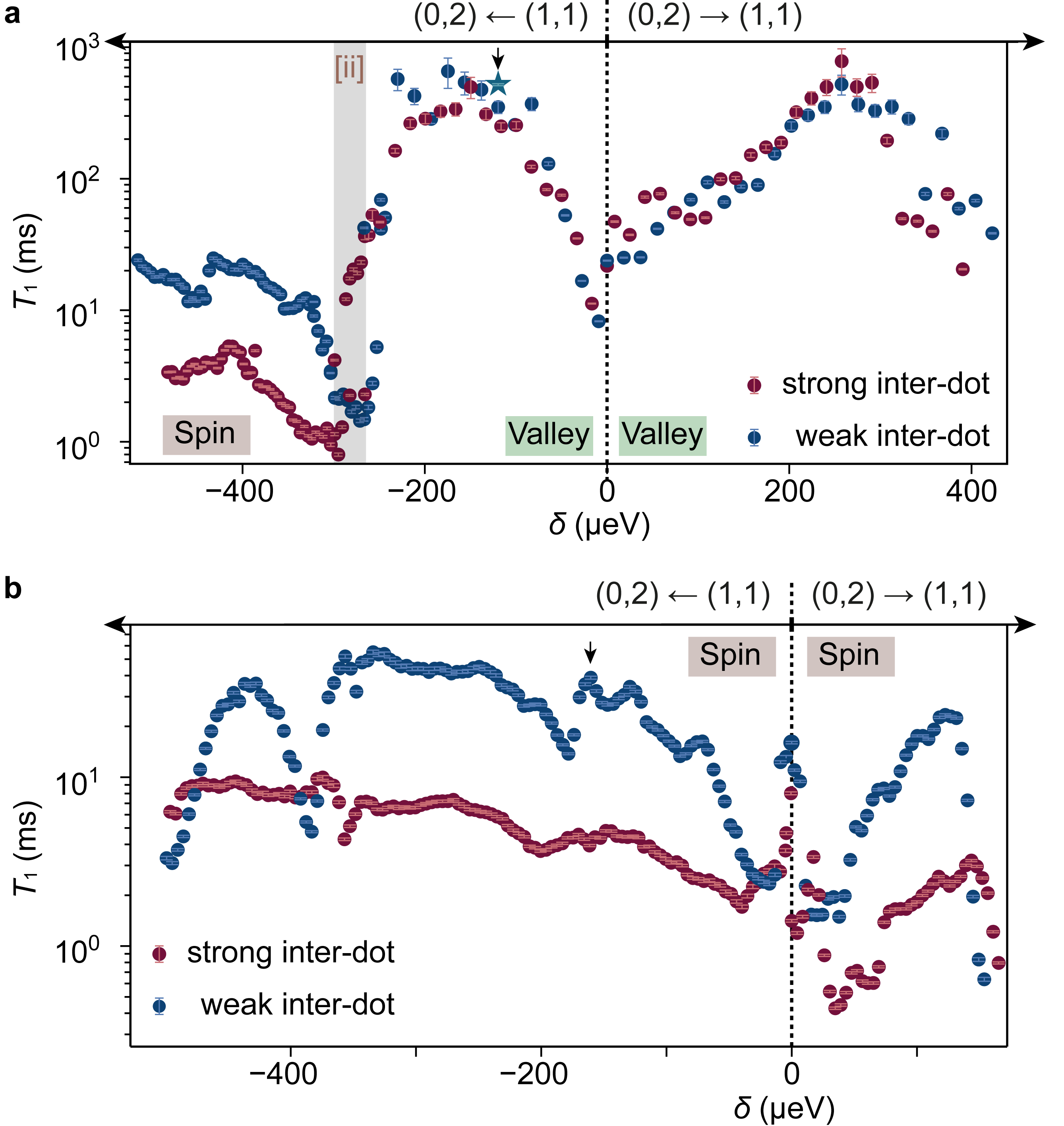}
	\caption{Spin and valley $T_1$ measured along the $\delta$-axis at $B_\mathrm{\perp}=\SI{250}{mT}$ (\textbf{a}) and $\SI{700}{mT}$ (\textbf{b}) for weaker (blue) and stronger (red) interdot tunnel couplings. Error bars correspond to the standard deviation of the calculated $T_1$. The star marks measurements with read time of $\SI{1.5}{s}$. 
    Spin- and valley-blockaded regions are labelled, with transition [ii] separating the two regions. When the electrochemical levels in the two dots are aligned, we observe a dip in $T_1$.}
	\label{figcom}
\end{figure}

For the quantitative measurement of $T_1$ times, we choose load locations L in the probability maps in Fig.~\ref{FIG2}a that only load the respective ground states while avoiding the excited states. We choose unload and load times, $t_\mathrm{U}$ and $t_\mathrm{L}$, longer than the dot--lead tunnelling rate, the time $t_\mathrm{A}$ spent at the anchor point $A$ longer than the rise time of our pulse lines at the order of $\SI{100}{\micro s}$, and read time $t_\mathrm{R}$ reasonably long compared to the probed relaxation times. Examples of time traces for $(1,1)\rightarrow(0,2)$ spin relaxation at $\SI{700}{mT}$ (at $\delta=\SI{-180}{\micro eV}$, marked by an arrow in Fig.~\ref{figcom}b) are shown in Fig.~\ref{figeg}a, left panel with $t_\mathrm{R}=\SI{100}{ms}$, where registered relaxation events are indicated by arrows. We show the distribution of relaxation times for \num{10000} repeated pulse cycles in Fig.~\ref{figeg}b, left panel, plotted on a logarithmic scale, which is well described by an exponential decay $e^{-t/T_1}$, with $T_1$ being the characteristic spin-relaxation time. To extract $T_1$, we perform a Bayesian analysis based on the exponential model $e^{-t/T_1}$ using the average relaxation time $\bkt{t}$ within the read-time interval $[t_0,t_1]$ as the relevant statistics. The finite read-time interval removes events occurring before the detection bandwidth $t_0$ and after $t_1$, the end of the read-window with the detection bandwidth subtracted (see Methods for a detailed explanation of the analysis performed). The data is well fitted with an exponential decay with $T_1=34\pm\SI{1}{ms}$, as extracted by this method.

The long-lived valley states are found for $(1,1)\rightarrow(0,2)$ by measuring valley relaxation at $B_\mathrm{\perp}=\SI{250}{mT}$ (at $\delta=\SI{-100}{\micro eV}$, marked by an arrow in Fig.~\ref{figcom}a) with read time $t_\mathrm{R}=\SI{1.5}{s}$, much longer than that for spin relaxation, to capture most relaxation events within the readout window. Examples of time traces are shown in Fig.~\ref{figeg}a, right panel, with the distribution of \num{10000} repeated pulse cycles plotted in Fig.~\ref{figeg}b, right panel. The valley-relaxation data are also well described by an exponential decay, allowing to extract a characteristic valley-relaxation time of $T_1=354\pm\SI{5}{ms}$ in this particular example.

We now evaluate quantitatively the valley and spin readout fidelities in our experiment. We prepare (1,1)$\mathrm{T^-}$ with a probability of roughly $50\%$ at the beginning of the read phase. Well-separated peaks corresponding to $(1,1)$ and $(0,2)$ charge states are seen when plotting histograms of detector voltage $V_\mathrm{det}$ during the read phase, as shown in Fig.~\ref{figeg}c and e. We follow the framework introduced in Barthel~et~al.~\cite{Barthel2009} to model the distribution which includes the effect of a finite relaxation time $T_1$, and find an overall fidelity of $\SI{99.9727(17)}{\percent}$ for the valleys. The shoulder in the lower histogram peak in Fig.~\ref{figeg}c is a result of charge instabilities close to the detector influencing its asymmetry sensitivity and thus shifting both the spin singlet and triplet levels with respect to zero, but by a different amount (see Supplementary Information~D). The charge instability could be avoided by more accurate tuning. We calculate the model function for each singlet peak using the mean and variance of a Gaussian fit 
to extract lower bounds for the fidelity of $99.9033(18)\%$ (solid) and $99.80(3)\%$ (dashed) for the spins. We plot the evolution of the infidelity with the length of the considered read phase in Fig.~\ref{figeg}d and find that our signal-to-noise ratio does not need improvement with higher statistics, but that the finite relaxation time $T_1$ limits the fidelity which decreases for longer $t_\mathrm{int}$ for both spins and valleys. The decrease of fidelity with $t_\mathrm{int}$ is faster for spins due to the shorter $T_1$.

The characteristic relaxation times $T_1$ of valley and spin states were measured as a function of detuning $\delta$ (see Fig.~\ref{fig1}b) by repeating the procedure described above. For valley relaxation, we choose a read time of $t_\mathrm{R}=\SI{140}{ms}$ for shorter measurements and compensated the shorter read time by larger statistics of \num{36000} pulse cycles instead of \num{10000} (see Methods for a more detailed discussion on this approach). The results are summarised in Fig.~\ref{figcom}. The blue star marks the single measurement with longer read time in the valley blockaded regime presented in Fig.~\ref{figeg}. It matches well with the $T_1$ extracted from shorter measurements. We see clearly a drop of the apparent $T_1$ time by roughly an order of magnitude as the detuning causes a change from valley to spin blockade at $\delta=-\SI{240}{\micro eV}$, for transition [ii] in Fig.~\ref{FIG2}a. In general, both spin and valley $T_1$ times show complex behaviour as a function of detuning. Our findings align with a similar trend of non-monotonic detuning dependence as highlighted in~\cite{johnson2005triplet}. It is important to note that the detuning range under scrutiny in our research is significantly offset from the (2,1) and (1,0) charge states. This implies that thermal relaxation may not be the primary factor at play. We attribute the dips in relaxation time partially to the coupling to excited states transitioning from the (1,1) triplet to the (0,2) charge state. These distinct peaks of increased relaxation, or "hotspots," correspond to situations of maximal overlap between these states. In such cases, phonon interactions facilitate the process, pushing the relaxation times towards a minimum. When moving away from these anticrossings, the relaxation time exhibits a recovery to its maximum values.

We also adjusted the strength of the tunnel coupling between the quantum dots, to probe its influence on the measured spin and valley relaxation rates. Any such dependence potentially contains information relevant for identifying the relaxation mechanisms in future work.
Figure \ref{figcom}a,b shows measured relaxation times at $B_\mathrm{\perp}=$ $\SI{250}{mT}$ and $\SI{700}{mT}$, respectively, as a function of detuning with $T_1$ times plotted on a logarithmic scale for two different voltages $V_\mathrm{MB}$ applied to the tunnel barrier gate between the two dots, resulting in stronger (red) and weaker (blue) interdot coupling strength, but both in the overall weak coupling regime. Spin- and valley-blockaded regions are labelled in accordance with the discussion of Fig.~\ref{FIG2}, separated by transition [ii].
We mark in Fig.~\ref{figcom} the locations of $\delta$ at which the examples in Fig.~\ref{figeg} are taken by arrows. 
We notice that the valley $T_1$ appears to be independent of interdot coupling, whereas the spin $T_1$ decreases consistently at both $\SI{250}{mT}$ and $\SI{700}{mT}$ by an order of magnitude from around $\SI{30}{ms}$ to $\SI{1}{ms}$ for stronger interdot coupling. This observation indicates that the mechanisms assisting spin relaxation are evidently dependent on interdot tunnelling, potentially hinting towards the involvement of momentum-dependent spin--orbit interactions~\cite{Stepanenko2012}. By contrast, for valley relaxation such mechanisms are clearly not the main contributors. Despite the shift of excited-state resonances due to the spin- and valley-state coupling to $B_\mathrm{\perp}$, no significant influence of $B_\mathrm{\perp}$ on the measured $T_1$ times can be concluded (for more details, see Supplementary Information~E).

\begin{figure}
	\includegraphics[width=8.5cm]{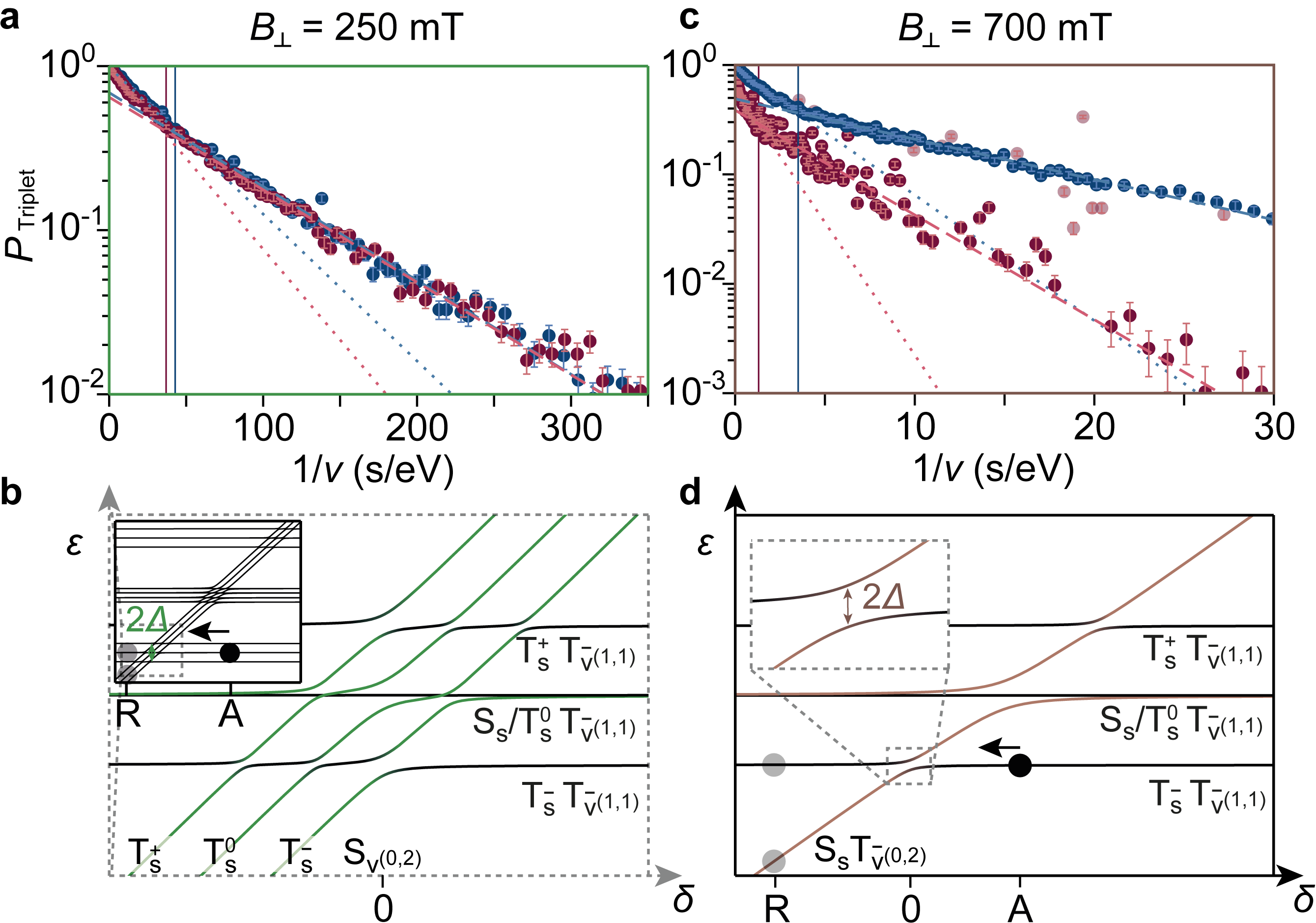}
	\caption{Probing the triplet $T^-$ singlet $S$ coupling $\Delta$ at \textbf{a} $B_\mathrm{\perp}=\SI{250}{mT}$ for valleys, and \textbf{c} $B_\mathrm{\perp}=\SI{700}{mT}$ for spins, using the Landau--Zener non-adiabatic transition probability $P_\mathrm{LZ}=e^{(-2\pi\Delta^2/v\hbar)}$, where $2\Delta$ is the minimum energy splitting between the states, and $v$ the energy sweep rate. Starting from point A with $(1,1)\mathrm{T^-}$, the probability of non-adiabatic retention of the $(1,1)$ state $P_\mathrm{triplet}$ is shown in \textbf{a,c} after passing through the anti-crossing at various energy sweep rate $v$, adjusted by fixing the read position and altering the pulse time. \num{2000} pulse cycles are performed for each sweep rate. Dual exponential decay behaviours with their transition marked by the vertical line are observed. The fitted $\Delta$ for \textbf{a} valley $\mathrm{T^-}$--$\mathrm{S}$ are $\SI{1.5}{neV}$ and $\SI{1.2}{neV}$ for weaker (blue), and $\SI{1.6}{neV}$ and $\SI{1.2}{neV}$ for stronger interdot coupling (red); analogously for \textbf{c} spin $\mathrm{T^-}$--$\mathrm{S}$ are $\SI{5.3}{neV}$ and $\SI{3.0}{neV}$ for weaker (blue), and $\SI{7.6}{neV}$ and $\SI{4.8}{neV}$ for stronger interdot coupling (red). The data points shown in transparent colouring are excluded from the fit. Error bars correspond to the standard deviation of the binomial probability density distribution. Strong interdot tunnel coupling increases spin $\mathrm{T^-}$--$\mathrm{S}$ coupling, but has little effect on the weaker valley $\mathrm{T^-}$--$\mathrm{S}$ coupling. \textbf{b,d} Energy diagrams of relevant $(1,1)$ and $(0,2)$ states.}
	\label{figLZ}
\end{figure}

We probe the coupling of resonant $(1,1)$ and $(0,2)$ states in the spirit of Landau--Zener tunnelling experiments.
In Fig.~\ref{FIG2}, peaks of probabilities are seen when states align in energy (marked by dashes), indicating finite coupling between the aligned states, which lifts the Pauli blockade.
For the data presented in Figs.~1--4, we have pulsed from the anchor point A to the read position R as fast as our line-bandwidth of the order of $\SI{10}{kHz}$ permits. 
In further experiments, we altered the transit time from A to R while keeping the read position R constant, thereby varying the energy sweep rate $v$. We repeated the procedure for \num{2000} pulse cycles at each sweep rate, and registered events where transfer from $(1,1)$ to $(0,2)$ happens diabatically. In Fig.~\ref{figLZ}a,c, the probability distribution $P_\mathrm{Triplet}$ of retaining $(1,1)\mathrm{T}$ after passing through the avoided crossing is plotted on a logarithmic scale against $1/v$ for the two magnetic fields a $B_\perp=\SI{250}{mT}$ and c $B_\perp=\SI{700}{mT}$, probing valley and spin blockade, respectively. The same experiment was performed for stronger (red) and weaker (blue) interdot tunnel coupling. In all cases, the data are compatible with the exponential dependence
$P_\mathrm{LZ}=\exp(-2\pi\Delta^2/v\hbar)$ predicted by the Landau--Zener formula, where $2\Delta$ is the minimum energy splitting between the states. In agreement with the data shown in Fig.~\ref{figcom}, intervalley coupling appears to be insensitive to the interdot tunnel coupling, whereas spin coupling clearly increases for stronger interdot tunnel coupling. 

For a quantitative analysis of the data, we look at the level schemes depicted in Fig.~\ref{figLZ}b,d. Here, the relevant $(1,1)$ triplet states marked in black, and the $(0,2)$ singlet states in green (valley singlets) or brown (spin singlets). At $B_\mathrm{\perp}=\SI{250}{mT}$, the initial state at point A belongs to a state of the $(1,1)\mathrm{T^-_v}$ bundle (see inset of Fig.~\ref{figLZ}b), while the resulting state at point $R$ is either in the same (1,1) bundle or in the (0,2) bundle of states (green in the main panel). The measured probability distribution $P_\mathrm{Triplet}$ is therefore a `bundle' distribution potentially giving the coupling between the two crossing state bundles of distinct valley character. By contrast, at $B_\perp=\SI{700}{mT}$, the distribution refers to the coupling between two distinct spin states, $(1,1)\mathrm{T_s^-T_v^-}$ and $(0,2)\mathrm{S_sT_v^-}$. We also note that all distributions in Fig.~\ref{figLZ}a,c seem to be double exponentials, with a fast decay rate at very small $1/v$ and a slower rate at larger $1/v$ (separated in the figure by vertical lines). To obtain a na\"ive estimate of the energy scales for the coupling of the involved states, we apply the Landau--Zener formula to both the fast and slow decay in all four traces. For the intervalley coupling $\Delta$ at $B_\perp=\SI{250}{mT}$ we find the values $\SI{1.5}{neV}$/$\SI{1.2}{neV}$ for weak, and $\SI{1.6}{neV}$/$\SI{1.2}{neV}$ for strong tunnel coupling (the two values correspond to the two slopes of the double exponential). At $B_\perp=\SI{700}{mT}$ for spin states we find $\SI{5.3}{neV}$/$\SI{3.0}{neV}$ for weak, and $\SI{7.6}{neV}$/$\SI{4.8}{neV}$ for strong interdot tunnel coupling. These values correspond to time scales $\hbar/\Delta$ of the order of a few hundred nanoseconds. For spins the same measurement technique~\cite{Shevchenko2010} gave a gap size of \SI{60}{neV} for gallium arsenide~\cite{Petta2010} and \SI{113}{neV} for silicon~\cite{Harvey-Collard2019}.

The double-exponential decay could arise from contributions of inelastic $T_1$ decay during the transit from point A to R, in addition to the coherent Landau--Zener physics accounted for by the transition probability $P_\mathrm{LZ}$. This scenario would tend to invalidate our na\"ive application of the Landau--Zener formula and make a more involved, possibly incoherent analysis necessary~\cite{Shimshoni1993,Krzywda2021}. Furthermore, the energy scales we extracted are extremely small, of the order of nanoelectronvolts, a factor of \num{1000} smaller than the temperature of the experiment. One might therefore expect that interactions of the electronic states with other degrees of freedom in the device become relevant, with virtual transitions, phonons or charge noise being among the most obvious candidates. We therefore regard the extracted coupling values as upper bounds of the true `intrinsic' values. 
Nonetheless, the experimental evidence of an exponential dependence on $1/v$ remains a robust outcome of our experiment.

Overall, the spin $T_1$ times of up to $\SI{60}{ms}$ in our experiments compare well with recent experiments on single quantum dots using the Elzerman readout \cite{GachterGarreis2022}.
The impressively long valley $T_1$ times of more than $\SI{500}{ms}$, which we show to be robust against interdot tunnel coupling, open up interesting avenues for exploiting valley physics, and together with the widely tunable valley $g$-factor~\cite{Tong2021} offer experimental schemes for electrically and coherently driven valley qubits.
At the same time our work raises the question how valley qubits can be manipulated by experimentally accessible parameters. The fact that the $K$-valleys are good quantum numbers relies on the translational invariance of the crystal. The electronic wave function in our dots extends over approximately $\SI{50}{nm}$, comprising some 400 lattice constants; this means that translational invariance remains a good concept. As layers get thinner, especially the insulating hBN layers, and gate geometries smaller, it is conceivable to create graphene quantum dots that are much smaller and tunable in size, possibly allowing gate manipulation of the valley degrees of freedom. A practical concept for operation and coherent control of valley qubits in graphene requires a better understanding of the mechanisms limiting the valley (and spin) $T_1$ times as observed in our experiments. The next experimental steps will include measurements with RF pulse lines, to observe the dephasing time $T_2$--- another crucial timescale for qubit operation---and of coherent valley oscillations in real time.

\section*{Acknowledgments}
We thank P. Märki and T. Bähler as well as the FIRST staff for their technical support. We thank A. Trabesinger for his valuable input during the writing process of this manuscript. K.E. acknowledges funding from the Core3 European Graphene Flagship Project, the Swiss National Science Foundation via NCCR Quantum Science and Technology, grant number FQXi-IAF19-07 from the Foundational Questions Institute Fund, ERC Synergy QUANTROPY No 951541, the European Union Horizon 2020 programme under grant agreement number 862660/QUANTUM E LEAPS, and the EU Spin-Nano RTN network. R.G. acknowledges funding from the European Union Horizon 2020 programme under the Marie Sk{\l}odowska-Curie grant agreement No 766025. K.W. and T.T. acknowledge support from JSPS KAKENHI (Grant Numbers 19H05790, 20H00354 and 21H05233).

\section*{Author contribution}
R.G. and C.T. contributed equally to this work. R.G. fabricated the sample. C.T. and R.G. performed the experiment with the help of W.W.H.. R.G., C.T., and J.T. analysed the data with the assistance of W.W.H. and J.D.G.. M.J.R. wrote the code for the pulse generation and readout with the lock-in amplifier. The pulsed measurements were set up by R.G., C.T., M.J.R., L.M.G., and W.W.H.. K.W. and T.T. synthesized the hBN crystals. K.E. and T.I. supervised the project. All authors discussed the results. R.G. and C.T. wrote the manuscript. All authors contributed to editing the manuscript.

\section*{Competing interests}
The authors declare no competing interests.


%

\section*{Methods}
\subsection*{Sample geometry}
The same device has been used before for the measurement of spin-relaxation times in Ref.~\onlinecite{GachterGarreis2022} and the evaluation of the full counting statistics in Ref.~\onlinecite{Garreis2022}. The fabrication of the van der Waals heterostructure follows the general procedure described in previous publications \cite{Hiske2018electrostatically, Eich2018, Banszerus2018}. Stacked with the dry-transfer technique \cite{wang2013drytransferedge}, it lies on a silicon chip with \SI{280}{nm} surface SiO$_2$. From bottom to top it is built up with a graphite back gate, a bottom hBN flake ($\SI{31}{nm}$), and a bilayer graphene flake capped with a top hBN flake ($\SI{20}{nm}$). The split gates ($\SI{5}{nm}$ Cr and $\SI{20}{nm}$ Au) are designed to form two channels with a nominal width of $\SI{100}{nm}$, with a separation gate of \SI{150}{nm} width in between them. The finger gates ($\SI{5}{nm}$ Cr and $\SI{20}{nm}$ Au) have a width of $\SI{20}{nm}$ and a centre-to-centre distance of $\SI{85}{nm}$. We use an aluminium oxide layer ($\SI{30}{nm}$) to separate the finger-gate layer from the split gates. Figure~\ref{fig1}a shows a false- colour atomic force microscope image of the two layers of metal gates fabricated on top of the heterostructure. The split gates (dark grey) are used to form two conducting channels (black) \cite{Hiske2018electrostatically}. For the measurements discussed in this paper we use two finger gates (blue) to define a quantum dot based on a p--n junction~\cite{Eich2018,Banszerus2018} in the lower channel, which we utilise as a charge detector~\cite{Kurzmann2019Detector}. In the second channel we define two quantum dots below the gates marked in dark-red colour and use the neighbouring gates (light red) to tune the tunnel coupling of the quantum dots to the leads as well as the interdot coupling~\cite{Tong2021, Banszerus2020}. All other gates are grounded.

\subsection*{Experimental set-up}
The sample is mounted in a dilution refrigerator with a nominal base temperature of $\SI{10}{mK}$; in previous measurements, we extracted an electronic temperature of $\SI{50}{mK}$ in the same device and set-up~\cite{Garreis2022}. The detector dot is biased with a constant current of $\SI{10}{pA}$ using a low noise differential amplifier~\cite{Marki2017} and the voltage signal $V_\mathrm{det}$ is measured with a detector bandwidth of about $\SI{1}{kHz}$ and sampled with a rate of $\SI{27.5}{kHz}$.\\
The dc voltage $V_\mathrm{L(R)}$ tuning the left (right) dot is combined with the pulse $V_\mathrm{pulse, L(R)}$ via two resistors at room temperature. The pulse lines to the sample have a rise time of $\SI{120}{\micro s}$. The pulse sampling rate to generate the analogue pulse signal is $\SI{219.72}{kHz}$.

\subsection*{Determination of $T_1$}

The finite memory of our arbitrary waveform generator limits our maximum read time to $t_\mathrm{R}=\SI{140}{ms}$. For the measurement presented in Fig.~\ref{figeg}b, we chose a point in detuning, where we can set the read position to a position corresponding to applying $V_\mathrm{pulse, L/R}=0$, i.e., at the centre of the pulse window. This allows us to turn off the arbitrary waveform generator at the end of the pulse sequence and reach an arbitrarily long read phase. This approach is only possible for data points within a small range of detuning, as else setting the centre of the pulse window at the read position means that the load and unload positions fall out of the pulse window. With this method, we chose a read phase of $t_\mathrm{R}=\SI{1.5}{s}$, much longer than the extracted $T_1 = 354 \pm \SI{5}{ms}$ at this point. This allows us to confirm the exponential distribution of relaxation events.

For any other data presented here, the length $t_\mathrm{R}$ of the read phase is not necessarily much longer than the relaxation time $T_1$, which means that we cannot estimate $T_1 = \langle t \rangle$ to describe the exponential decay. Instead, we use Bayes' theorem to find the posterior distribution function of $\gamma = 1/T_1$ given the data and evaluate its maximum for estimating $T_1$ and its width for estimating the uncertainty of $T_1$.\\
We select all time traces that show a transition in the time interval $[t_\mathrm{min},t_\mathrm{max}]$, where $t_\mathrm{min} = \SI{1}{ms}$ and $t_\mathrm{max} = t_\mathrm{R} - \SI{1}{ms}$. The time interval of $\SI{1}{ms}$ corresponds to the detector rise time. All the remaining traces are discarded, as they are either traces with initial decays and hence wrong initialisation, or traces without a decay within $t_\mathrm{R}$.
The model distribution is then
\begin{equation}
\pdf(t) = \frac{1}{\alpha(\gamma)}e^{-\gamma t}\gamma
\label{dist1}
\end{equation}
with $\gamma>0$. The normalisation constant is given by the condition
\[ \int_{t_\mathrm{min}}^{t_\mathrm{max}} \pdf(t)\,dt = \frac{1}{\alpha(\gamma)}\int_{t_\mathrm{min}}^{t_\mathrm{max}} e^{-\gamma t}\gamma dt = 1\]
and therefore
\[ \alpha(\gamma) = \int_{t_\mathrm{min}}^{t_\mathrm{max}} e^{-\gamma t}\gamma dt = e^{-\gamma t_\mathrm{min}}-e^{-\gamma t_\mathrm{max}}. \]

From the sequence of $N$ experimental time traces, we obtain a sequence of decay-time data of the form
\[ D = \{ t_1,t_2,t_3,\ldots, t_{N} \}, \]
where $N$ is the number of traces that showed a decay between $t_\mathrm{min}$ and $t_\mathrm{max}$. The probability to measure this specific data set, if $\gamma$ is known (the likelihood of the dataset $D$) is
\begin{multline*}
\prob(D|\gamma) = \prod_{i=1}^{N}\left(\frac{1}{\alpha(\gamma)}e^{-\gamma t_i}\gamma dt_i\right)\\
= \frac{\gamma^N}{\alpha(\gamma)^N}e^{-\gamma\sum_{i=1}^{N}t_i}\prod_{i=1}^{N}dt_\mathrm{i}.
\end{multline*}
We now introduce the time average
\[ \langle t_i\rangle = \frac{1}{N}\sum_{i=1}^{N}t_i. \]
This allows us to write
\[ \prob(D|\gamma) = \frac{\gamma^N}{\alpha(\gamma)^N}e^{-\gamma N \langle t_i\rangle}\prod_{i=1}^{N}dt_\mathrm{i}.\]
Using Bayes' theorem, we find the posterior distribution of $\gamma$, given the a specific dataset $D$:
\[ \pdf(\gamma | D) = \frac{\pdf(\gamma)\prob(D | \gamma)}{\int_0^\infty \pdf(\gamma)\prob(D | \gamma)d\gamma}. \]
A suitable non-informative prior for the scaling variable $\gamma$ is
\[ \prob(\gamma) = \frac{d\gamma}{\gamma}. \]
This leads us to
\begin{equation}
\pdf(\gamma | \langle t_i\rangle, N)= \frac{\frac{\gamma^{N-1}}{\alpha(\gamma)^N}e^{-\gamma N\langle t_i\rangle}}{\int_0^\infty \frac{\gamma^{N-1}}{\alpha(\gamma)^N}e^{-\gamma N\langle t_i\rangle}d\gamma}.
\label{posterior}
\end{equation}
This is a distribution function for $\gamma$ with a sharp peak. The maximum of this distribution function gives the most probable value for $\gamma$, and its width the associated uncertainty.

The denominator in the posterior distribution is a constant.
The numerator is a function of $\gamma$, which we define to be
\[ h(\gamma) = \frac{\gamma^{N-1}}{\alpha(\gamma)^N}e^{-\gamma N\langle t_i\rangle}. \]
Finding the maximum of the posterior probability density function in eq.~\eqref{posterior} amounts to finding the maximum of $h(\gamma)$. Numerically, this is most conveniently done by realising that the maximum of $h(\gamma)$ is also the maximum of $\ln(h(\gamma))$. We find
\[ \ln(h(\gamma)) = (N-1)\ln\gamma - N\ln(\alpha(\gamma))-\gamma N\langle t_i\rangle\,. \]
The maximum of this function is found by solving
\[ \frac{d\ln(h(\gamma))}{d\gamma} = \frac{N-1}{\gamma} - \frac{N}{\alpha(\gamma)}\frac{d\alpha(\gamma)}{d\gamma} - N\langle t_i\rangle = 0,\]
so that
\[ \frac{1}{\gamma} = \frac{N}{N-1}\left(\langle t_i\rangle + \frac{1}{\alpha(\gamma)}\frac{d\alpha(\gamma)}{d\gamma}\right) .\]

\section*{Data availability}
The data supporting the findings of this study are made available via the ETH Research Collection~\cite{datarepository}.

\clearpage


\section*{Supplementary Information}

\setcounter{figure}{0} 
\renewcommand\thefigure{A\arabic{figure}}

\section{A.~Blockade of (0,2) and (1,1) states}

\begin{figure}
	\includegraphics[width=8.5cm]{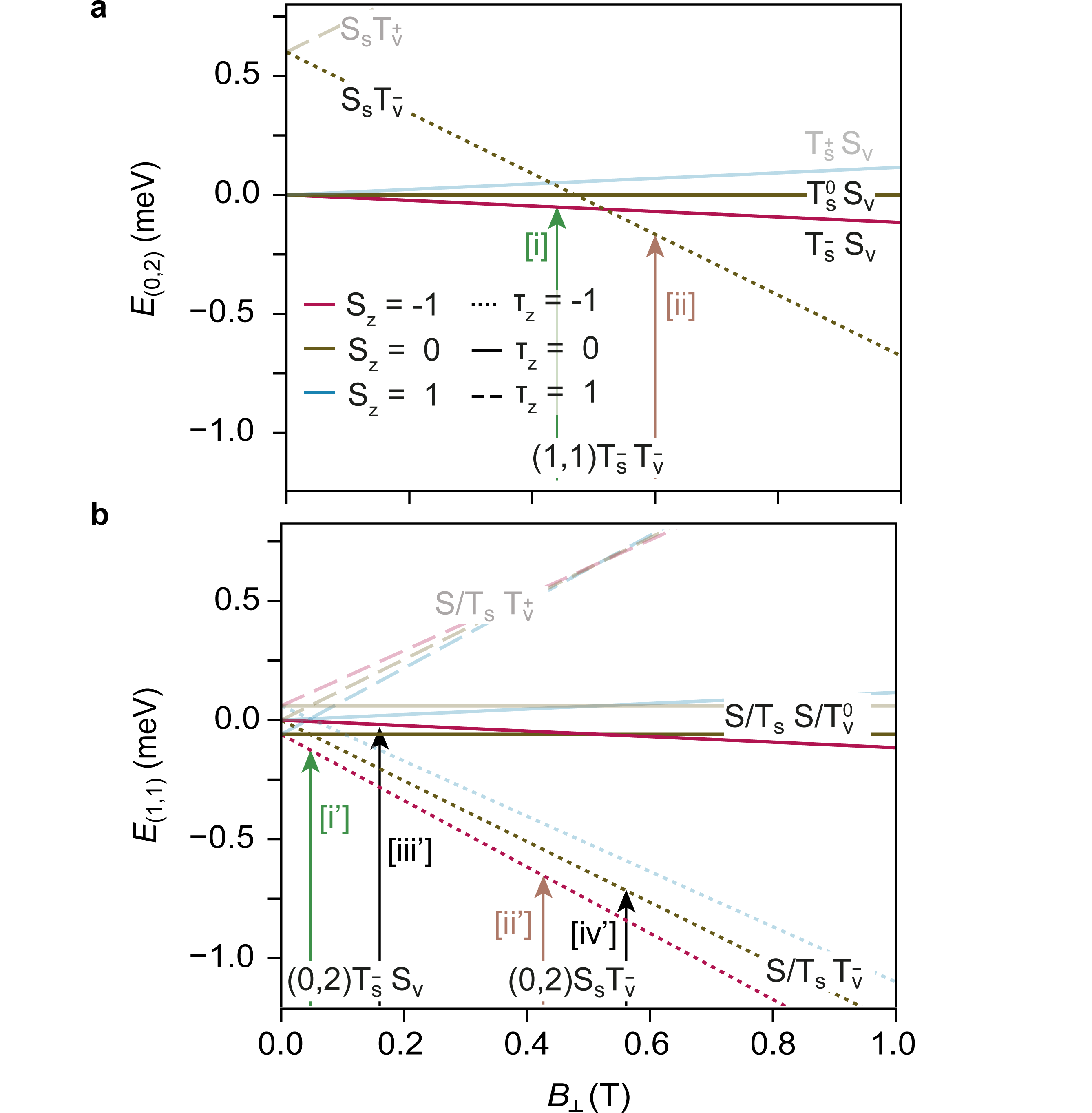}
	\caption{Energies of \textbf{a} $(0,2)$ and \textbf{b} $(1,1)$ states in dependence of $B_\mathrm{\perp}$, sketched for spin $g$-factor $g_\mathrm{s}=2$ and valley $g$-factor $g_\mathrm{v}=22$, Kane--Mele zero-field splitting $\Delta_\mathrm{SO}=\SI{60}{\micro eV}$, and $(0,2)$ singlet-triplet splitting $E_\mathrm{ex}=\SI{600}{\micro eV}$. The two-particle ground states switch from valley-singlet to spin-singlet state at around \SI{600}{mT}. Accordingly, the ground-state blockade switches from valley to spin blockade.}
	\label{figA1}
\end{figure}

The nature of Pauli blockade depends on the states involved, which can be switched by an external perpendicular magnetic field~\cite{Kurzmann219excited, Tong2022}.
A detailed comparison of theoretical calculations~\cite{Knothe2020} and experiment can be found in literature~\cite{Moeller2021}. For convenience, we plot the relevant (0,2) and (1,1) states in Fig.~\ref{figA1}, where we represent the respective spin and valley number by colour and line style, respectively. As the valley $g$-factor is much larger than the spin $g$-factor, the (0,2) ground state changes at around \SI{450}{mT} from a spin-triplet valley-singlet to a spin-singlet valley-triplet. For lower fields, the (1,1) and (0,2) ground-state valley numbers do not match (valley blockade), while for higher fields the spin quantum numbers differ (spin blockade). For finite detuning, the excited states can change the nature of blockade.

\section{B.~Double quantum dot charge stability map}
\begin{figure}
	\includegraphics[width=8.5cm]{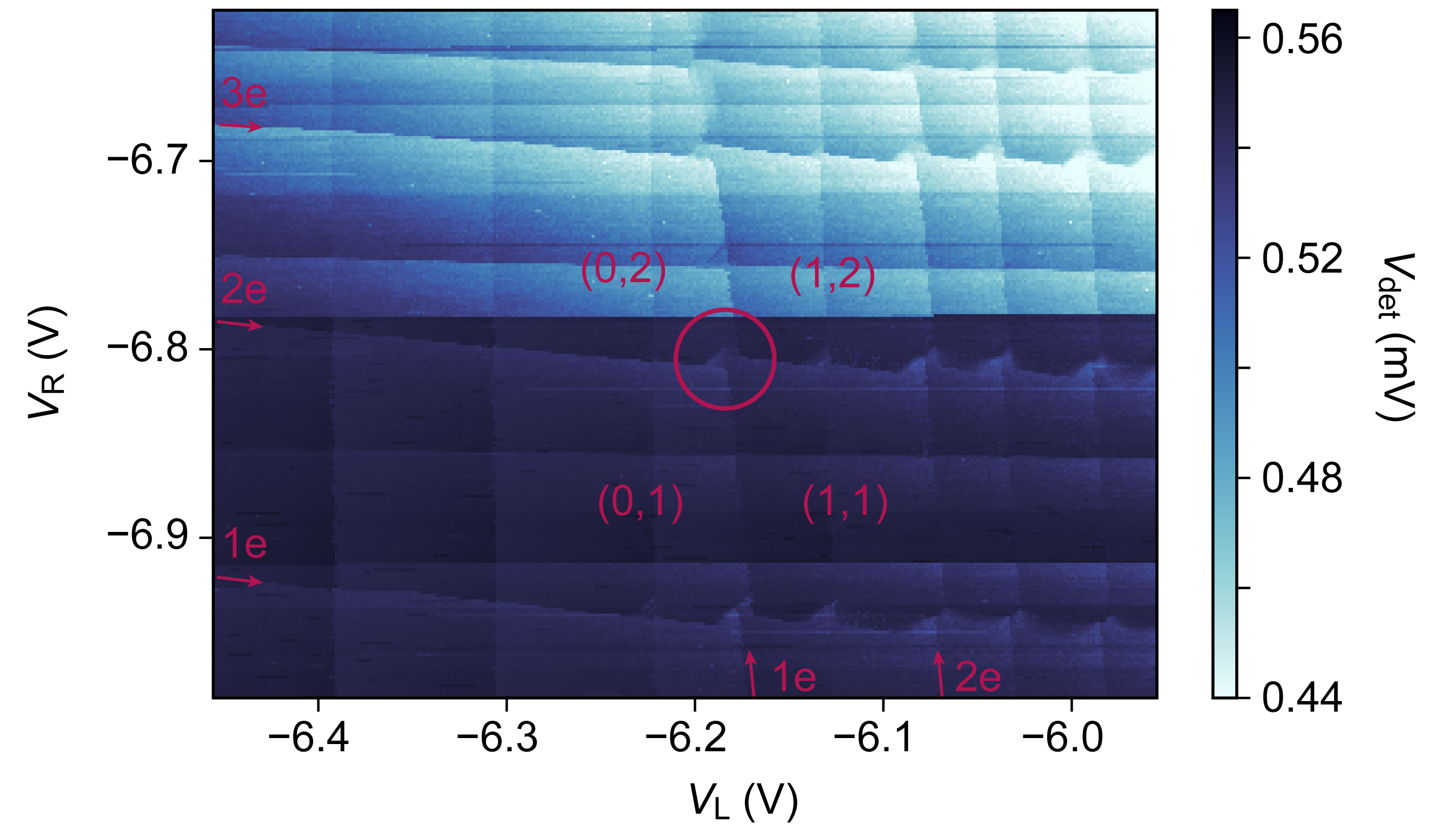}
	\caption{Double quantum dot charge stability map for a larger range of plunger gate voltages, at bias voltage $V_\mathrm{SD}=\SI{1}{mV}$. The relevant charge transition lines for the electron dot L and dot R are labelled by red arrows. In this experiment, we operate at the circled triple points. The additional resonances arise from dots formed under the dot-lead barrier gates.}
	\label{figAy}
\end{figure}

For orientation, we plot the double quantum dot charge stability map for a larger range of plunger gate voltages and finite bias voltage $V_\mathrm{SD}=\SI{1}{mV}$ in Fig.~\ref{figAy}. The relevant charge transition lines for the electron dot L and dot R are labelled by red arrows. In this experiment, we operate at the circled triple points. The additional resonances with a different slope arise from dots formed under the dot-lead barrier gates.

\section{C.~Pulsed charge stability maps in $B_\perp$}

\begin{figure}
	\includegraphics[width=8.5cm]{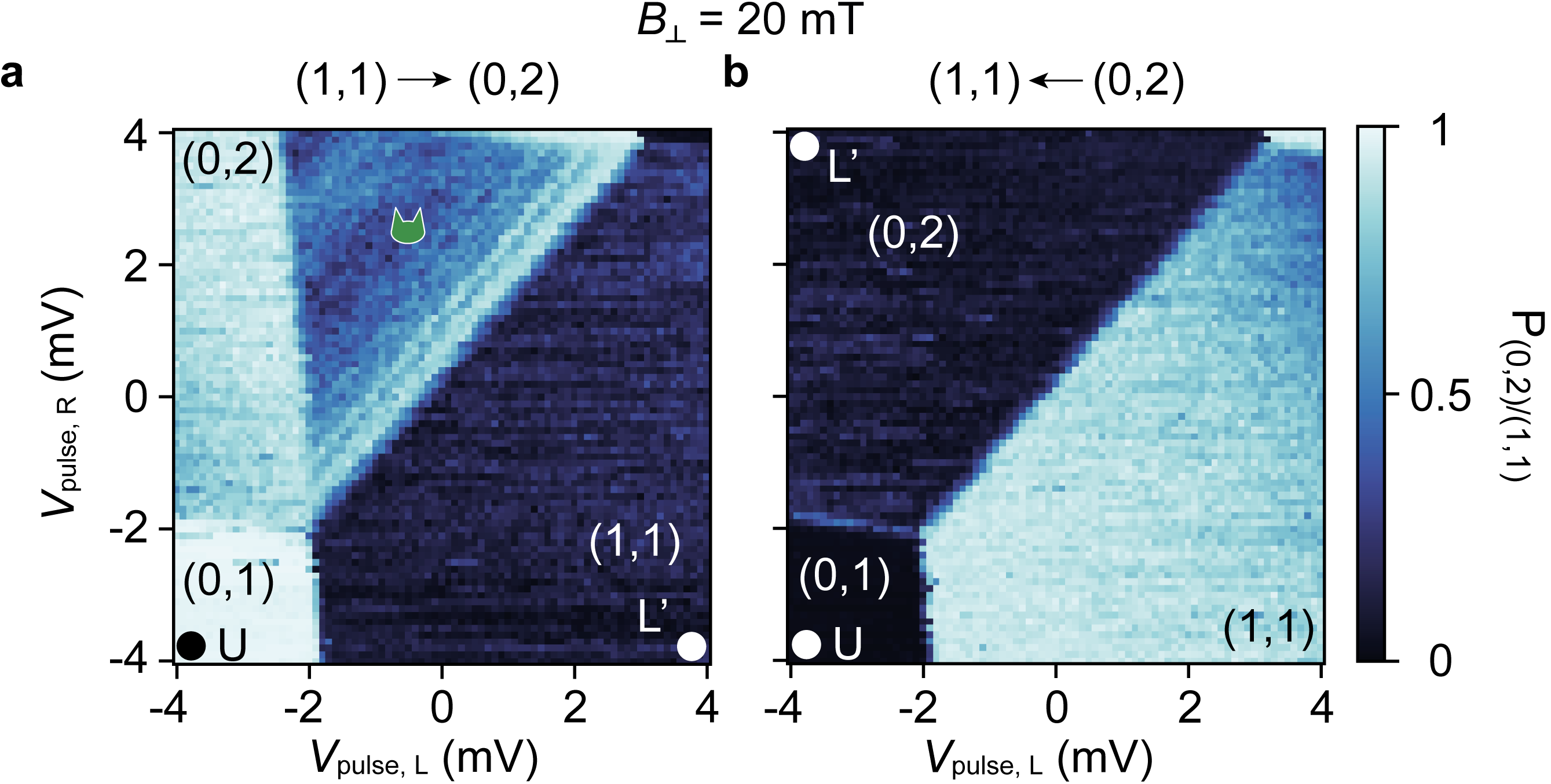}
	\caption{Probability of occupation at $B_\perp = \SI{20}{mT}$ in \textbf{a} $(0,2)$ and \textbf{b} $(1,1)$ states during the read phase for pulsing \textbf{a} $(1,1)\rightarrow(0,2)$ and \textbf{b} $(0,2)\rightarrow(1,1)$ while loading at locations L' ($B_\mathrm{\perp}=\SI{20}{mT}$). The green dot labels the valley-blockade region, shown by the low occupation probability (dark) for the target state.}
	\label{figA2}
\end{figure}

As shown in Fig.~\ref{figA2}, at $B_\mathrm{\perp}=\SI{20}{mT}$, we observe on pulsing from $(1,1)$ to $(0,2)$ a darker triangular region (green dot), within which the system has non-zero probability to be in $(1,1)$ during the read phase, even though $(0,2)$ is energetically preferred. The single-dot two-electron $(0,2)$ ground state is a spin-triplet valley-singlet, denoted $(0,2)\mathrm{T_sS_v}$ (see Fig.~\ref{figA1}), whereas the loaded $(1,1)$ state could be of any spin or valley character. If a $(1,1)\mathrm{T^-_{v}}$ is loaded, the system will remain in $(1,1)$ due to Pauli valley blockade, even if $(0,2)\mathrm{S_{v}}$ is lower in energy. By contrast, no such blockaded region is observed for $(0,2)\rightarrow(1,1)$, as any loaded $(0,2)$ can transition into $(1,1)$ due to the abundance of energetically close $(1,1)$ states. At $B_\mathrm{\perp}=\SI{20}{mT}$ there would be a valley-blockaded region with a range of around $\SI{25}{\micro eV}$, narrower than the resonance width at $\delta = 0$ and therefore not resolvable. The same argument holds for the spin-blockaded state which is yet another $\Delta_\mathrm{SO} \sim \SI{60}{\micro eV}$ away.

\begin{figure}
	\includegraphics[width=8.5cm]{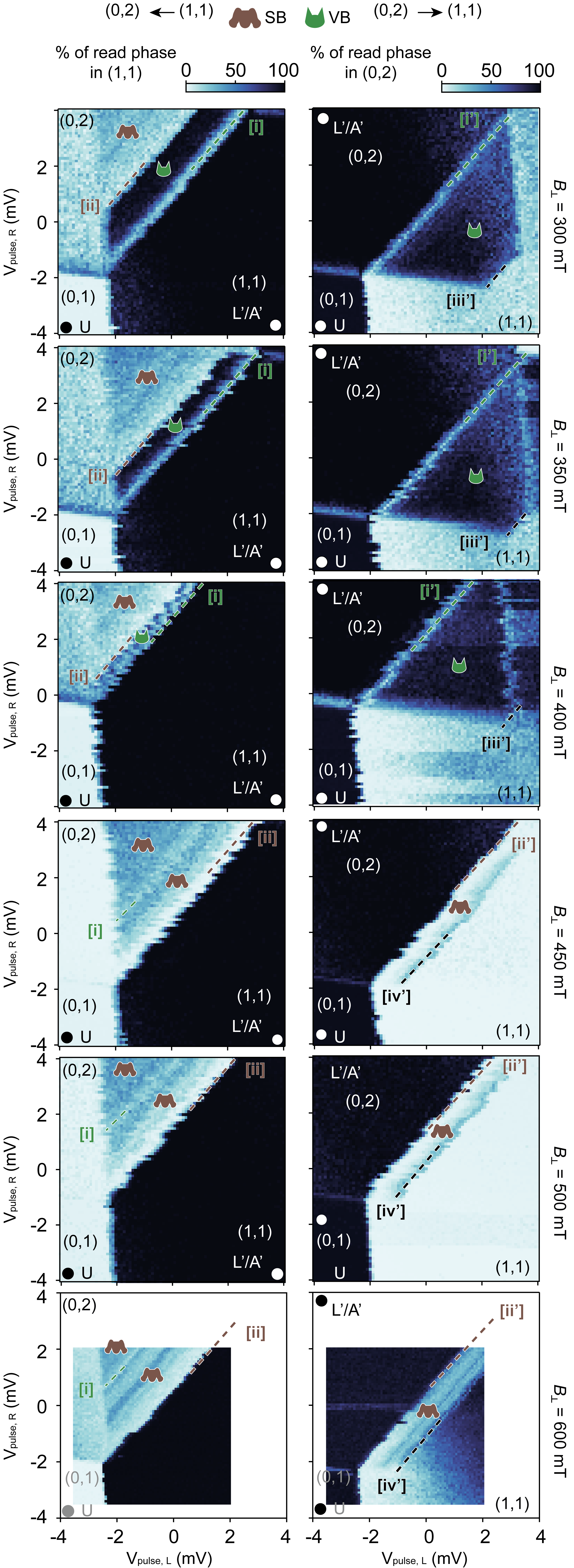}
	\caption{Probability of occupation in left $(0,2)$ and right $(1,1)$ states during the read phase for pulsing left $(1,1)\rightarrow(0,2)$ and right $(0,2)\rightarrow(1,1)$ while loading and anchoring at locations L'/A'. The green (brown) dot labels the valley- (spin-) blockade region, shown by the lower occupation probability (darker) for the target state.}
	\label{figAx}
\end{figure}

We present the evolution of pulsed charge stability maps for $B_\perp$ at $\SI{300}{mT}$, $\SI{350}{mT}$, $\SI{400}{mT}$, $\SI{450}{mT}$, $\SI{500}{mT}$, and $\SI{600}{mT}$. The change of the (0,2) ground state from a spin-triplet valley-singlet to a valley-triplet spin-singlet occurs at around $\SI{425}{mT}$. At $\SI{400}{mT}$ we reach a maximum region of valley blockade in $(0,2)\to(1,1)$ direction, while in the $(1,1)\to(0,2)$ direction the very small valley blocked region indicates that the two (0,2) ground states are now very close in energy. 

After the change of ground state, in $(0,2)\to(1,1)$ the large valley blockade region abruptly disappears, and we are left with a small region of spin blockade. The blockade observed for the $(1,1)\to(0,2)$ direction also abruptly changes from a very small region of valley to a large region of spin blockade. The movement of the excited states in $B_\perp$ agrees well with our understanding of the two-particle state spectra.

\section{D.~Details on the measurement fidelity}

\begin{figure}
    \includegraphics[width=8.5cm]{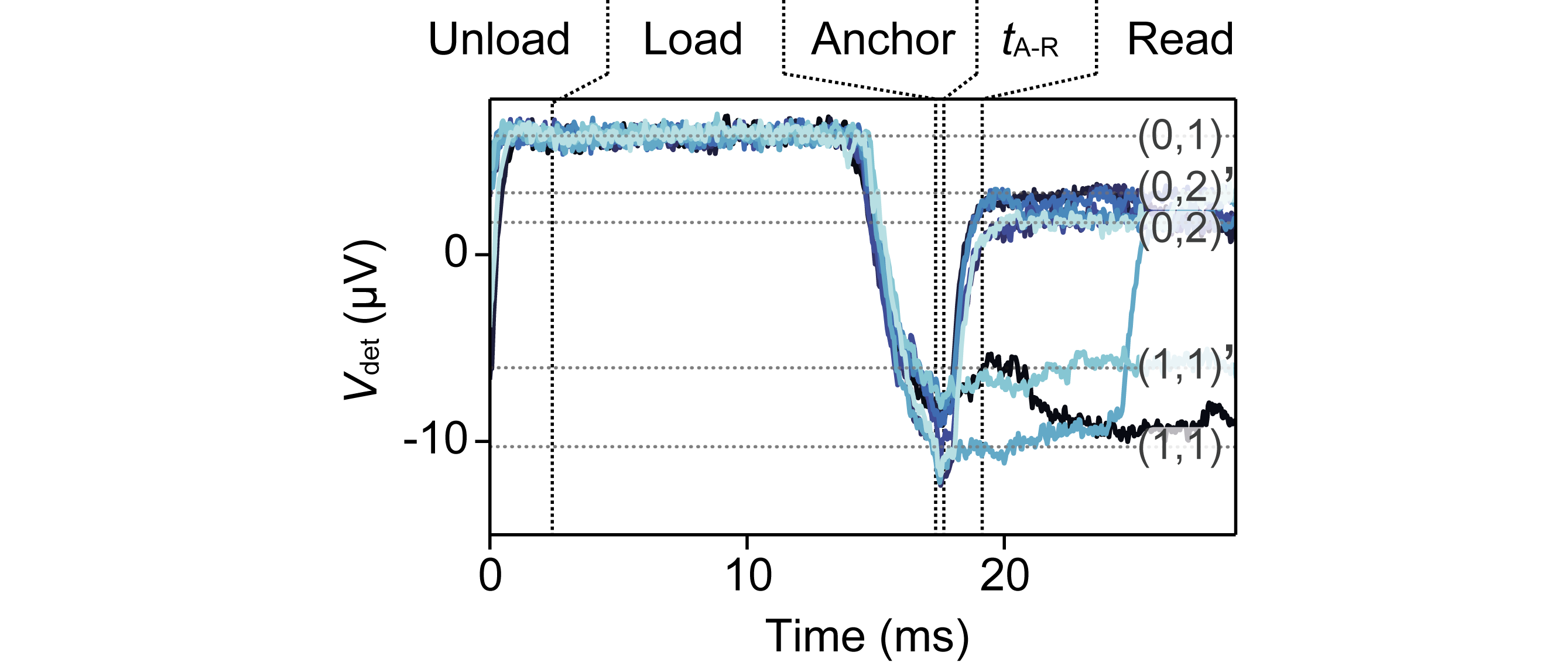}
    \caption{Exemplary time traces in the spin blockaded regime for a finite ramp time $t_\mathrm{A-R} = \SI{1.54}{ms}$ from A to R, corresponding to the histogram presented in Fig.~3c. The detector is sensitive to an additional but much slower charge rearrangement close by, shifting the detection level of the (1,1) and (0,2) states between two discrete levels.}
    \label{fig:exemplarytracesspin}
\end{figure}

By implementing a ramp time $t_\mathrm{A-R}$ between the anchor and read point, we can tune how adiabatically the gap $\Delta$ between $(0,2)S_sT^-_v$ and $(1,1)T^-_sT^-_v$ is crossed and hence vary the occupation probability of the respective $(1,1)$ and $(0,2)$ charge states as long as the sweep rate is faster than the relaxation time. We plot the histograms of the detector voltage $V_\mathrm{det}$ during the read phase in Fig.~3c,e for an occupation probability of roughly $50\%$. To understand the shoulder in the lower histogram peak in Fig.~3c, we plot exemplary time traces in Fig.~\ref{fig:exemplarytracesspin}. We see that the detector is sensitive to slow charge instabilities close by, influencing its asymmetry sensitivity between the two quantum dots and thus shifting both the $(1,1)$ and $(0,2)$ level with respect to $(0,1)$; but by a different amount. In any case, the two levels of different charge configuration are well separated such that this effect does not influence the analysis provided in this manuscript, since single pulses are evaluated independently with a local threshold. We reasonable assume that the two (1,1) levels are decoupled and can hence be treated as two separate Gaussian contributions to the histogram. Fitting the distribution for the spins with three and for the valleys with two Gaussians~\cite{Singh2018} $A \exp\left[-(x-\mu)^2/2\sigma^2\right]$, we can extract a signal to noise ratio of 4.8 (6.0) for spins (valleys). We evaluate individual readout fidelities for the singlet $F_S$ and triplet $F_T$ states by integrating the Gaussian fit over the respective histogram peak. We calculate the overall fidelity as $F = F_T + F_S - 1$, and we find a maximum of overall fidelity of $99.956(13)\%$ ($99.981(2)\%$) for spins (valleys).

\section{E.~$T_1$ dependence on magnetic field}

\begin{figure}
	\includegraphics[width=8.5cm]{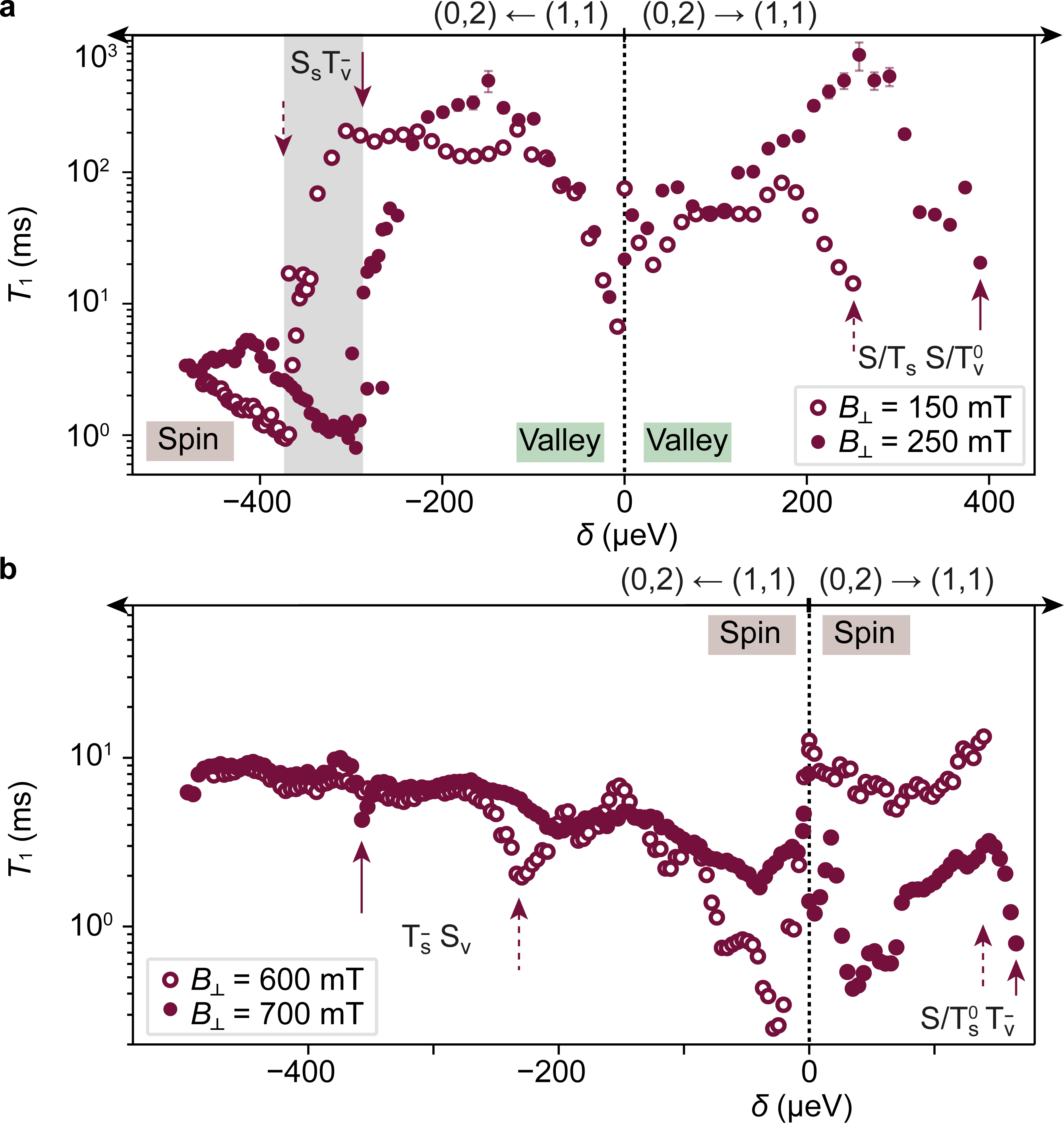}
	\caption{Spin (brown) and valley (green) $T_1$ measured along the $\delta$-axis at \textbf{a)} below and \textbf{b)} above the ground-state transition, for the same magnetic field $B_\perp$ as shown in Fig.~4 (solid) and \SI{100}{mT} lower (empty). We observe a shift of the resonances with magnetic field, but no obvious change in the overall $T_1$.}
	\label{figA3}
\end{figure}

In order to further characterise the nature of spin and valley blockade in our system, we perform a similar line cut along the detuning axis as shown in Fig.~4 for magnetic fields that were \SI{100}{mT} lower, that is, for $B_\perp = \SI{150}{mT}$ (Fig.~\ref{figA3}a) and $B_\perp = \SI{600}{mT}$ (Fig.~\ref{figA3}b). We observe no obvious field dependence for neither the spin nor the valley relaxation times. Due to the finite coupling between the (1,1) and (0,2) states, an alignment of the starting ground state with an excited state in the target charge configuration leads to a dip in $T_1$ if the nature of the dominating blockade mechanism is the same. We mark the respective excited states in Fig.~\ref{figA3}, which shift by $g\mu_\mathrm{B}B$ with $g$ either $g_v$ or $g_s$ and $\mu_\mathrm{B}$ the Bohr magneton. Any other dips and peaks in $T_1$ cannot be attributed to level crossings and might be the result of a specific phonon density of states.

%

\end{document}